\documentclass[aps,prb,reprint,nofootinbib,groupedaddress,citeautoscript,amsmath]{revtex4-2}
\usepackage{graphicx}
\usepackage{latexsym,amssymb,amsmath,bm,bbm}
\usepackage[caption=false]{subfig}
\usepackage{xcolor}
\usepackage{mathrsfs}
\usepackage[colorlinks=true,
urlcolor=blue,
linkcolor=blue,
citecolor=blue
]{hyperref}
\usepackage{esvect}
\usepackage[normalem]{ulem}

\renewcommand{\Im}{{\rm Im}}
\newcommand{\Tr}{{\rm Tr}}
\begin{document}
\title{Topological phonon polariton enhanced radiative heat transfer in bichromatic nanoparticle arrays mimicking Aubry-Andr\'e-Harper model}

\author{B. X. Wang}
%\email{wangboxiang@sjtu.edu.cn}
\affiliation{Institute of Engineering Thermophysics, School of Mechanical Engineering, Shanghai Jiao Tong University, Shanghai 200240, China}
\affiliation{MOE Key Laboratory for Power Machinery and Engineering, Shanghai Jiao Tong University, Shanghai 200240, China}
%\affiliation{Key Laboratory of Thermal Management and Energy Utilization of Aircraft, Ministry of Industry and Information Technology, Nanjing 210016, China}
\author{C. Y. Zhao}
\email{changying.zhao@sjtu.edu.cn}
\affiliation{Institute of Engineering Thermophysics, School of Mechanical Engineering, Shanghai Jiao Tong University, Shanghai 200240, China}
\affiliation{MOE Key Laboratory for Power Machinery and Engineering, Shanghai Jiao Tong University, Shanghai 200240, China}
\date{\today}

\begin{abstract}
Topological phonon polaritons (TPhPs) are promising optical modes relevant in long-range radiative heat transfer, information processing and infrared sensing, whose topological protection is expected to enable their robust existence and transport. In this work we show that TPhPs can be supported in one-dimensional (1D) bichromatic silicon carbide nanoparticle (NP) chains, and demonstrate that they can considerably enhance radiative heat transfer for an array much longer than the wavelength of radiation. By introducing incommensurate or commensurate modulations on the interparticle distances, the NP chain can be regarded as an extension of the off-diagonal Aubry-Andr\'e-Harper (AAH) model. By calculating the eigenstate spectra with respect to the modulation phase that creates a synthetic dimension, we demonstrate that under this type of modulation the chain supports nontrivial topological modes localized over the boundaries, since the present system inherits the topological property of two-dimensional integer quantum Hall systems. In this circumstance the gap-labeling theorem and corresponding Chern number can be used to characterize the features of band gaps and topological edge modes. Based on many-body radiative heat transfer theory for a set of dipoles, we theoretically show the presence of topological gaps and midgap TPhPs can substantially enhance radiative heat transfer for an array much longer than the wavelength of radiation. We show how the modulation phase that acts as the synthetic dimension can tailor the radiative heat transfer rate by inducing or annihilating topological modes. We also discuss the role of dissipation in the enhancement of radiative heat transfer. These findings therefore provide a fascinating route for tailoring near-field radiative heat transfer based on the concept of topological physics.   
\end{abstract}

\maketitle
\section{Introduction} 
The discovery of topological phases of matter has led to a plenty of research interests in novel topological quantum materials \cite{xiaoNaturereviewphys2021} and inspired fascinating analogies in photonic \cite{ozawa2018topological}, atomic \cite{atalaNaturephys2013}, acoustic \cite{heNaturephys2016} and mechanical \cite{susstrunkScience2015} systems. The most unique feature of these topological systems is that they can support strongly localized and unidirectionally propagating edge or interface states which are largely immune against the presence of disorder and impurities, thanks to the topological protection. Topological protection means that the existence of these states is a global property regarding the topology of the entire band structure, which disappears only if the gap closes \cite{hasanRMP2010}. As one of the most promising analogies, topological photonic systems \cite{khanikaevNaturemat2013,luNPhoton2014,khanikaevNPhoton2017} not only provide a playground for observing topological modes directly and exploring novel physics that is not easily accessible in electronic systems like long-range interactions \cite{liNaturephoton2019} and non-Hermitian topology \cite{zhaoScience2019}, but also have promising applications in unidirectional waveguides \cite{poliNComms2015}, optical isolators \cite{el-GanainyOL2015,KarkiPRApplied2019}, topological lasers \cite{partoPRL2018,zengNature2020,shaoNaturenano2020} and topological sensors \cite{wangPRM2020} and so on. 

A seemingly separate research field is thermal radiation heat transfer. Recently, Biehs and coworkers \cite{ottPRB2020,ottPRB2021,ottIJHMT2022} and the authors \cite{wangPRB2018b,wangJAP2020} proposed the possibility that topological optical modes may play an important role in the control of thermal radiation, especially in tailoring near-field radiative heat transfer. In fact, topological photonic systems generally reply on the specific arrangements of different micro/nanoscale elements, which naturally bear an intrinsic connection with typical systems regarding many-body radiative heat transfer \cite{benabdallahPRL2011,dongPRB2017,chenJQSRT2018,fangPRB2020,biehsRMP2021,songRRP2021}, if a temperature gradient exists within them. The most typical case is nanoparticle (NP) arrays. For instance, Biehs and coworkers \cite{ottPRB2020,ottPRB2021,ottIJHMT2022} recently investigated plasmonic InSb NP arrays that are specifically arranged to mimic the one-dimensional (1D) and two-dimensional (2D) Su-Schrieffer-Heeger (SSH) models as well as the quantum spin Hall (QSH) system in honeycomb lattices. They showed that in the topologically nontrivial phases, thermal near-field energy density is significantly enhanced at the edges and corners (if there are topological corner states), and the edge modes opened additional heat flux channels, which dominate the radiative heat transport.

In this work, we attempt to study the topological edge states and radiative heat transfer in a similar system with richer underlying physics, that is a bichromatic NP array mimicking the well-known Aubry-Andr\'e-Harper (AAH) model \cite{krausPRL2012,krausPRL2012b,ganeshanPRL2013,verbinPRL2013,poshakinskiyPRL2014}. This model is a 1D tight-binding lattice model with on-site or/and hopping terms being cosine modulated. When this cosine modulation is incommensurate (commensurate) with the lattice, this system becomes quasiperiodic (periodic). In this manner, this type of lattice model is sometimes dubbed ``bichromatic" \cite{alpeggianiOptica2019}. The presence of cosine modulation gives rise to nontrivial topological properties that can be mapped to the 2D integer quantum Hall (IQH) system (namely, the well-known Harper-Hofstadter model in a square lattice with a perpendicular magnetic field), in the absence of a realistic magnetic field \cite{ozawa2018topological,krausPRL2012}. In this situation, the modulation phase $\phi$ plays the role of momentum (wavenumber) in a perpendicular synthetic dimension, which leads to a dimensional extension to 2D \cite{krausPRL2012,krausPRL2012b,chalopinNaturephys2020}.  As a consequence, this model provides a playground for studying profound topological phase transitions and topological states in 1D. 

Different from previous works regarding radiative heat transfer in topological many-particle systems regarding plasmonic NPs \cite{ottPRB2020,ottPRB2021,ottIJHMT2022}, our interest here is focused on the topological phonon polaritons (TPhPs), since they not only inherit the properties of deep-subwavelength confinement and low loss of phonon polaritons that are prominent for mediating significant near-field radiative heat transfer \cite{liScience2018,foteinopoulouNanophoton2019,volokitinRMP2007,songNaturenano2015,dongPRB2018,chenIJHMT2021,zhangIJHMT2022,fangIJHMT2022}, but also exhibit topological protection and strong localization over the edges \cite{wangPRB2018b}. These features render TPhPs promising in long-range radiative heat transfer, information processing and infrared sensing, in which topological protection is expected to enable their robust existence and transport.

In this work, we show that TPhPs can be supported in 1D bichromatic silicon carbide nanoparticle chains by introducing incommensurate or commensurate modulations on the interparticle distances, as an extension of the off-diagonal AAH model. We calculate the band structures (eigenstate spectra) with respect to the modulation phase $\phi$, which plays the role of a synthetic dimension. We find the evidence showing the present system inherits the topological properties of 2D IQH systems, and the spectral position and number of these topologically protected edge modes are governed by the gap-labeling theorem, which dictates the topological invariant, i.e., the Chern number, indicating the validity of bulk-boundary correspondence. Based on many-body radiative heat transfer theory for a set of dipoles, we theoretically show the presence of topological gaps and mid-gap phonon polariton edge modes can considerably enhance radiative heat transfer, for an array much longer than the wavelength of thermal radiation. We show how the modulation phase that acts as the synthetic dimension can tailor the radiative heat transfer rate by inducing or annihilating topological modes. We also discuss the role of dissipation of the SiC material. These findings therefore provide a fascinating route for tailoring near-field radiative heat transfer based on the concept of topological physics.

\section{Model}
Consider a 1D array composed of spherical $\alpha-$ (hexagonal) SiC NPs aligned along the $x$-axis. To mimic the AAH model, we introduce artificial modulations over the spacings between adjacent NPs, given by \cite{poshakinskiyPRL2014,pilozziPRB2016}
\begin{equation}\label{modulation}
x_{n+1}-x_{n}=d[1+\eta\cos(2\pi \beta n+\phi)],
\end{equation}
where $x_n$ denotes the position of the $n$-th NP, $d$ introduces the on-average interparticle distance (or the periodic lattice constant before modulation), $\eta$ determines the amplitude of the distance modulation, $\beta$ is the periodicity of the modulation that controls the bichromaticity of the lattice. If it is irrational, we say the modulation is incommensurate while if it is rational the modulation is then commensurate. And $\phi$ stands for the modulation phase that corresponds to the momentum (wavenumber) in a synthetic orthogonal dimension \cite{krausPRL2012,krausPRL2012b,ganeshanPRL2013,verbinPRL2013,poshakinskiyPRL2014}.

\subsection{The electric dipole approximation and coupled-dipole model}
Since SiC NPs support strongly localized phonon polariton resonances in the infrared region around $11\mathrm{\mu m}$ due to excitation of longitudinal optical phonons, the permittivity of SiC can be described by a Lorentz model as \cite{wheelerPRB2009}
\begin{equation}\label{permittivity}  \varepsilon_p(\omega)=\varepsilon_\infty\Big(1+\frac{\omega_L^2-\omega_T^2}{\omega_T^2-\omega^2-i\omega\gamma}\Big),
\end{equation}
in which $\omega$ is the angular frequency of the driving field in the unit of $\mathrm{cm}^{-1}$ (wavenumber), $\varepsilon_\infty=6.7$ is the high-frequency limit of permittivity, $\omega_T=790 \mathrm{cm}^{-1}$ is the angular frequency of transverse optical phonons,
$\omega_L=966 \mathrm{cm}^{-1}$ is the angular frequency of longitudinal optical phonons, and $\gamma=5\mathrm{cm}^{-1}$ is the non-radiative damping coefficient \cite{wheelerPRB2009}. Without loss of generality, the radius of the spherical SiC NP is fixed as $a=0.1\mathrm{\mu m}$. Such a small NP (much smaller than the resonance wavelength of localized phonon polaritons) can be approximated as an electric dipole \cite{wangPRB2018b}. In this situation the electromagnetic response of an individual SiC NP can be described by the dipole polarizability within the so-called radiative correction to balance scattering and extinction \cite{tervoPRMater2018,markelPRB2007,parkPRB2004}:
\begin{equation}\label{radiativecorrection}
	\alpha(\omega)=\frac{4\pi a^3\alpha_0}{1-2i\alpha_0(ka)^3/3}
\end{equation}
with 
\begin{equation}
	\alpha_0(\omega)=\frac{\varepsilon_p(\omega)-1}{\varepsilon_p(\omega)+2}.
\end{equation}

%The extinction efficiency calculated under the ED approximation using the polarizability with radiative correction is also shown in Fig.\ref{qextsingle}, and a good agreement with the exact Mie theory is observed. An extra examination of the validity of radiative correction for different particle sizes is given in Appendix \ref{rad_correct_append}.

When the distance between the centers of adjacent spherical NPs is less than $3a$, the EM response of the entire array is described by the well-known set of coupled-dipole equations \cite{parkPRB2004,markelPRB2007,tervoPRMater2018}:
\begin{equation}\label{coupled_dipole_eq}
	\mathbf{p}_j(\omega)=\alpha(\omega)\left[\mathbf{E}_\mathrm{inc}(\mathbf{r}_j)+\frac{\omega^2}{c^2}\sum_{i=1,i\neq j}^{\infty}\mathbf{G}_0(\omega,\mathbf{r}_j,\mathbf{r}_i)\mathbf{p}_i(\omega)\right]
\end{equation}
for $j=1,2,...,N$, where $c$ is the speed of light in vacuum. $\mathbf{E}_\mathrm{inc}(\mathbf{r})$ is the external incident field (if any) and $\mathbf{p}_j(\omega)$ is the excited electric dipole moment of the $j$-th NP. $\mathbf{G}_{0}(\omega,\mathbf{r}_j,\mathbf{r}_i)$ is the free-space dyadic Green's function describing the propagation of field emitting from the $i$-th NP to $j$-th NP, which is give by \cite{markelPRB2007}
\begin{equation}
	\begin{split}
\mathbf{G}_{0}(\omega,\mathbf{r}_i,\mathbf{r}_j)&=\frac{\exp{(ikr)}}{4\pi r}\Big[\left(\frac{i}{kr}-\frac{1}{k^2r^2}+1\right)\mathbf{I}\\&+\left(-\frac{3i}{kr}+\frac{3}{k^2r^2}-1\right)\mathbf{\hat{r}}\mathbf{\hat{r}}\Big]
	\end{split}
\end{equation} 
with $r=|\mathbf{r}|=|\mathbf{r}_i-\mathbf{r}_j|>0$ being the distance between two NPs and $\mathbf{\hat{r}}$ being the unit vector with respect to $\mathbf{r}$.

For 1D arrays, there are two types of electromagnetic modes, including the transverse and longitudinal ones \cite{weberPRB2004}. For the longitudinal modes, the dipole moments of the NPs are aligned to the $x$-axis, and therefore only the $xx$-component of the Green's function (GF) need to be used in calculation:
\begin{equation}\label{Gxx}
	G_{0,xx}(x)=-2\Big[\frac{i}{k|x|}-\frac{1}{(k|x|)^2}\Big]\frac{\exp{(ik|x|)}}{4\pi |x|},
\end{equation}
where $k=\omega/c$ is the free-space wavenumber. On the other hand, for transverse eigenstates whose dipole moments are perpendicular to the array axis, the transverse ($yy$ or $zz$) component of the GF is then used:
\begin{equation}\label{Gyy}
	G_{0,yy}(x)=[\frac{i}{k|x|}-\frac{1}{(k|x|)^2}+1]\frac{\exp{(ik|x|)}}{4\pi |x|},
\end{equation}

To determine the band structure (eigenstate distribution) of a finite NP array, we can set the incident field in Eq. (\ref{coupled_dipole_eq}) to be zero \cite{weberPRB2004,pocockArxiv2017}. Then an eigenvalue equation can be obtained:
\begin{equation}
\mathbf{M}|\mathbf{p}\rangle=\alpha^{-1}(\omega)|\mathbf{p}\rangle. 
\end{equation}
Here $\mathbf{M}$ stands for the interaction matrix whose elements are derived from the GF [Eqs. (\ref{Gxx}) and (\ref{Gyy}) according to the polarization of eigenmodes], for instance, for longitudinal modes, $M_{ij}=(\omega^2/c^2)G_{0,xx}(x_i-x_j)$ for $i\neq j$ and $M_{jj}=0$. The right eigenvector of this equation $|\mathbf{p}\rangle=[p_1p_2...p_j...p_N]$, in the braket notation, stands for the dipole moment distribution of an eigenstate, where $p_j$ is the dipole moment of the $j$-th NP. This equation can be solved to get a series of complex eigenfrequencies in the lower complex plane in the form of $\tilde{\omega}=\omega-i\Gamma/2$, where the real part $\omega$ amounts to the angular frequency of the eigenstate while the imaginary part $\Gamma$ corresponds to its linewidth (or decay rate of the eigenstate) \cite{caoRMP2015,pocockArxiv2017}. 

Since the topologically protected eigenstates are highly localized over the boundary of the finite chain \cite{atalaNaturephys2013}, to recognize these eigenstates clearly, we use the inverse participation ratio (IPR) to quantitatively measure for the localization degree of an eigenstate \cite{wangOL2018,wang2018topological}: 
\begin{equation}
	\mathrm{IPR}=\frac{\sum_{j=1}^{N}|p_j|^4}{(\sum_{j=1}^{N}|p_j|^2)^2}.
\end{equation}
An eigenstate with $\mathrm{IPR}=1$ is completely localized while an $\mathrm{IPR}=1/n$, where $n$ is an integer, indicates the eigenstate can be regarded as evenly distributed over $n$ NPs \cite{wangOL2018}. Therefore, for a highly localized topological edge state, its IPR should be much larger compared to those of the bulk eigenstates \cite{wangOL2018}. But we should note not all eigenstates with high IPRs are topologically protected. For instance, Anderson localized states in the bulk are topologically trivial.

\subsection{Calculation of radiative heat flux}
Within the dipole approximation, the radiative heat transfer can be calculated semi-analytically by employing the fluctuation-dissipation theorem (FDT). Radiative heat transfer in such a system is the simplest case of many-body radiative heat transfer, which has been discussed extensively in recent works \cite{benabdallahPRL2011,dongPRB2017,chenJQSRT2018,fangPRB2020,biehsRMP2021,songRRP2021}. Let us start from the coupled dipole model with fluctuating dipoles due to thermal excitation, which is \cite{benabdallahPRL2011}
\begin{equation}\label{fluctuation_cdm}
\mathbf{E}_{ij}=\mu_{0}\omega^{2}\mathbf{G}_{0}^{ij}\mathbf{p}_{j\neq i}^{\rm fluc}+\frac{\omega^{2}}{c^{2}}\underset{k\neq i}{\sum}\mathbf{G}_{0}^{ik}\alpha_{k}\mathbf{E}_{kj},
\end{equation}
in which $\mathbf{G}_{0}^{ij}\equiv\mathbf{G}_{0}(\omega,\mathbf{r}_i,\mathbf{r}_j)$ for brevity, and $\mathbf{E}_{ij}$ is the exciting electric field impinging at particle $i$ generated by the thermally induced fluctuating dipole moment of particle $j$. The first term in the right hand side is the direct propagation of a dipole field from the source $\mathbf{p}_{j\neq i}^{\rm fluc}$, while the second summation term indicates the scattering processes from other particles $k$ to particle $i$ due to the exciting electric fields $\mathbf{E}_{kj}$ impinging on them. If we consider the total GF $\mathbf{G}^{ij}$ for the propagation of electromagnetic waves from the fluctuating dipole $j$ with inclusion of the many-body scattering processes, which satisfies
\begin{equation}
\mathbf{E}_{ij}=\mu_{0}\omega^{2}\mathbf{G}^{ij}\mathbf{p}_{j\neq i}^{\rm fluc}.
\end{equation}
It can be directly solved from Eq. (\ref{fluctuation_cdm}). Such total GF can be obtained for all particles with fluctuating dipoles. Then by invoking FDT, the heat transfer rate from particle $j$ to $i$ is given by
\begin{equation}\label{interparticle_heatrate}
\mathcal{P}_{j\rightarrow i}=3\int_{0}^{\infty}\frac{d\omega}{2\pi}\,\Theta(\omega,T_{j})\mathcal{T}_{i,j}(\omega)
\end{equation}
in which $\Theta(\omega,T)=\hbar\omega/[\exp{(\hbar\omega/(k_bT))}-1]$ is the Planck oscillator with $\hbar$ and $k_b$ being the Planck and Boltzmann constants respectively, and the transmission coefficient is defined as
\begin{equation}
	\mathcal{T}_{i,j}(\omega)=\frac{4}{3}\frac{\omega^{4}}{c^{4}}\chi_{i}\chi_{j}\Tr\bigl[\mathbf{G}^{ij}\mathbf{G}^{ij\dagger}\bigr],
\end{equation}
where $\chi_{i}=\Im(\alpha_i)-k_0^3|\alpha|^2/(6\pi)$. A detailed derivation and description of above formulas can refer to Refs. \cite{biehsRMP2021,benabdallahPRL2011}.

\section{Incommensurate lattice}
\begin{figure*}[ht]
	\centering
	\includegraphics[width=1\linewidth]{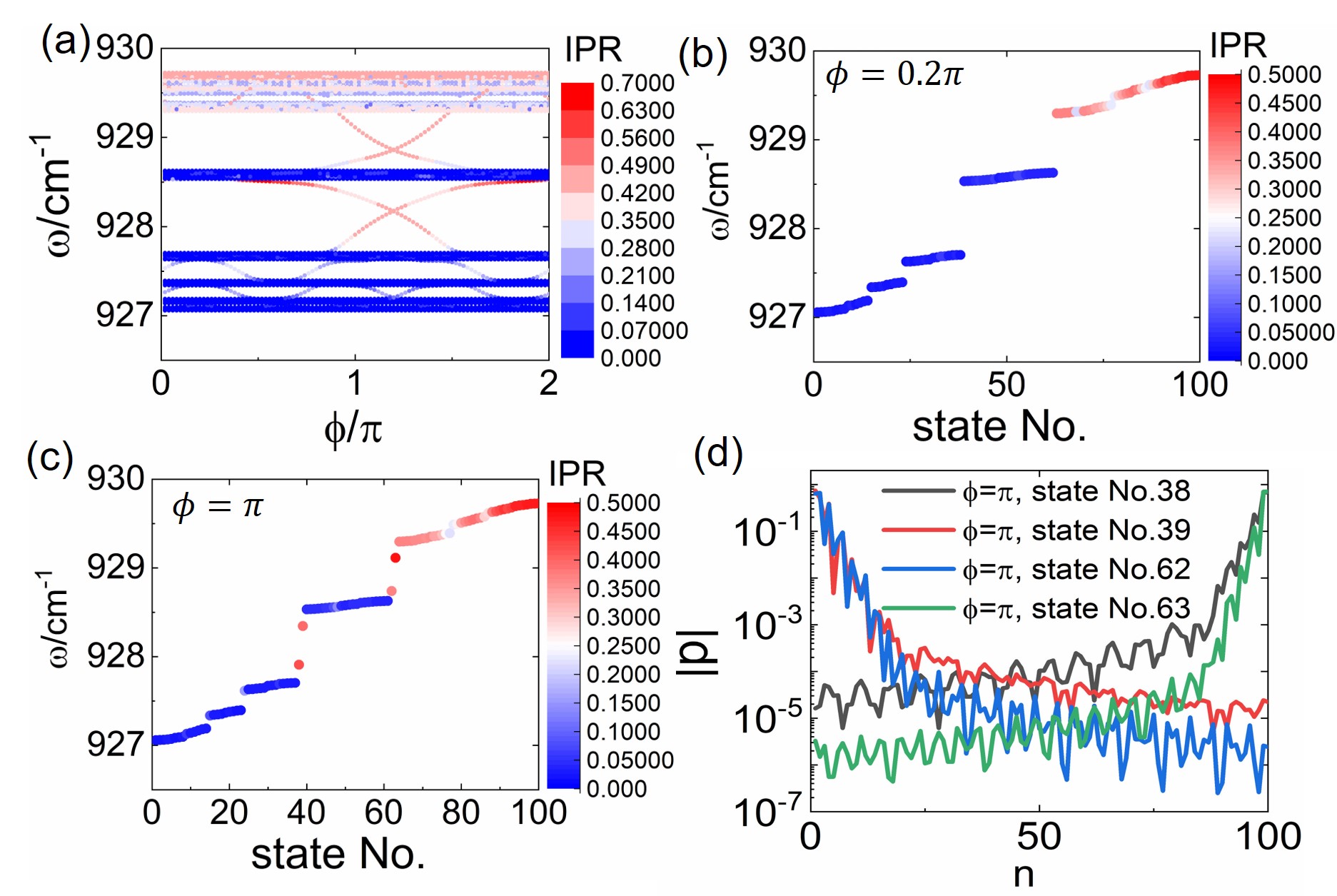}
	\caption{Longitudinal band structures of incommensurate lattices with $\beta=(\sqrt{5}-1)/2$, $d=0.6~\mathrm{\mu m}$, $\eta=0.3$ and $N=100$. (a) Band structure a function of modulation phase $\phi$. (b) Eigenstate spectrum for $\phi=0.2\pi$. (c) Eigenstate spectrum for $\phi=\pi$. (d) Dipole moment distributions of several typical midgap states.}
	\label{incommlongedgestates}
\end{figure*}

As mentioned, if $\beta$ is irrational, we say the modulation is incommensurate and the NP array becomes quasiperiodic. Quasiperiodic system is the intermediate phase with long-range order between periodic and fully disordered (random) systems, thus harboring a qualitatively different spectrum \cite{jagannathanRMP2021}. Here, we consider the case of $\beta=(\sqrt{5}-1)/2$, which is most commonly investigated incommensurate AAH lattice in previous works. We can plot the calculated eigenstate spectrum (band structure) as a function of $\phi$, which varies in the range from $0$ to $2\pi$. In Fig. \ref{incommlongedgestates}(a), the longitudinal band structure of a chain with $N=100$ NPs and $d=0.6~\mathrm{\mu m}$ is shown, in which the color of eigenstates stand for their IPRs. The modulation amplitude is chosen to be a moderate value of $\eta=0.3$, which does not affect of the generality of this work \cite{wangPRA2021}. It can be observed that the band structures break into a set of bands, in which large main gaps can be clearly seen with several discernible minigaps as well as many indiscernible gaps, that can be only seen in an enlarged figure (not shown here) \cite{wangPRA2021}. This is due to the irrational nature of the interparticle distance modulation that leads to a fractal spectrum. In fact, for an infinitely long chain, its spectrum constitutes a Cantor set with Lebesgue measure zero \cite{kohmotoPRL1983b,kohmotoPRL1987,avilaAnnalsMath2009,longhiPRB2019}). This is a typical feature of quasiperiodic systems. Moreover, all the bands are actually flat and the eigenfrequencies are almost unchanged with the variation of $\phi$ with a few states crossing the gaps, since the irrational modulation makes the system insensitive to lattice translation \cite{krausPRL2012b}.

In the two main gaps, namely, the gaps covering $927.7\mathrm{cm}^{-1}\lesssim\omega\lesssim928.5\mathrm{cm}^{-1}$ and $928.6\mathrm{cm}^{-1}\lesssim\omega\lesssim929.3\mathrm{cm}^{-1}$ in Fig. \ref{incommlongedgestates}(a), midgap states with high IPRs can be clearly recognized. These states are topologically protected edge states dictated by a nonzero Chern number, similar to the behavior of the conventional AAH model, as will be discussed below. These midgap states only exist for a specific range of modulation phase $\phi$'s. To see this more clearly, let us consider the eigenstate spectra of two cases, $\phi=0.2\pi$ [Fig. \ref{incommlongedgestates}(b)] and $\phi=\pi$ [Fig. \ref{incommlongedgestates}(c)]. The state number is assigned according to its frequency. In the former case, no midgap states can be found, where the high-IPR eigenstates are actually bulk states that are Anderson-localized \cite{wangPRA2021} (not shown here). In the latter case, in each of the two main gaps, a pair of midgap states are observed. The state numbers are denoted by 38, 39, 62 and 63 respectively. The dipole moment distributions over the chain for these states are given in Fig. \ref{incommlongedgestates}(d), showing they are strongly localized over the boundaries, as a manifestation of topological edge states. And in each of the main gaps, the pair of midgap states are localized over the left and right edges respectively. We further find that in each of the two main gaps, by varying the modulation phase $\phi$, the midgap edge states keep localized over the same edge as long as they remain in the gap. Therefore, these midgap states localized over the same edge in the same band gap can be regarded as belonging to the same ``mode" as $\phi$ plays the role of momentum in an extended dimension. For each of the two main gaps, there are two edge modes traversing the spectral gap, one localized over the left edge and the other localized over the right edge. This property is a manifestation of the topological nature of the band gaps \cite{krausPRL2012b}.

The topology property of the conventional AAH model can be characterized by the gap Chern number $\nu$, which satisfies the following equation  \cite{thoulessPRL1982,macdonaldPRB1984,danaJPC1985,amitPRB2018}:
\begin{equation} \label{gaplabel_eq}
\mathcal{N}=\mu+\nu\beta,
\end{equation}
in which $\mu$ is an integer and $\mathcal{N}$ is the normalized integrated density of states (IDOS) in the gap. This equation is a general result derived from the magnetic translational symmetry in an IQH system with 2D Bloch electrons subjected to rational (hence described by a rational $\beta=p/q$ with $p,q$ denoting  two coprime integers) magnetic fields. By taking the irrational limit for $\beta$, it can also be applied to incommensurate systems \cite{danaJPC1985,danaPRB1985,danaPRB2014}. More details on the origin of this equation can refer to Refs. \cite{simonAAM1982,luckPRB1989,bellissardRMP1992,danaPRB2014,wangPRA2021,jagannathanRMP2021}. This equation has only one set of solution $(\mu,\nu)$ for an irrational $\beta$ and a fixed $\mathcal{N}$, and therefore any band gaps with the same $\mathcal{N}$ and irrational $\beta$ can be labeled by the same set of integers $(\mu,\nu)$, independent of system details. As a consequence this equation is called the gap-labeling theorem \cite{simonAAM1982,luckPRB1989,bellissardRMP1992,danaPRB2014,wangPRA2021,jagannathanRMP2021}. This universality of these topological integers under an irrational $\beta$ is robustly protected by the magnetic translational symmetry \cite{cookmeyerPRB2020}. On the other hand, the situation is quite different for rational $\beta$'s, where there are infinite solutions for the equation and therefore the topological property is system-dependent \cite{thoulessPRL1982,macdonaldPRB1984,hatsugaiPRB1990,amitPRB2018}.

To use the gap-labeling theorem, the normalized IDOS $\mathcal{N}$ of a band gap can be calculated, which is the number of eigenstates below the gap divided by the total number of eigenstates in the spectrum for a specific $\phi$. In Fig. \ref{incommlongedgestates}(a), the lower main gap ($927.7\mathrm{cm}^{-1}\lesssim\omega\lesssim928.5\mathrm{cm}^{-1}$) has an IDOS of $\mathcal{N}\approx38/100=0.38$, which leads to a solution for the gap-labeling theorem of $\mu=1,\nu=-1$. The upper main gap $928.6\mathrm{cm}^{-1}\lesssim\omega\lesssim929.3\mathrm{cm}^{-1}$ exhibits an IDOS $\mathcal{N}\approx62/100=0.62$, resulting in an integer solution of $\mu=0$, $\nu=1$. Therefore, the Chern number of the lower (upper) main gap, as a global property from the bulk band structure, is $\nu=-1$ ($\nu=+1$). According to the bulk-boundary correspondence in 2D IQHE, for a gap Chern number of $\nu$, there must be $|\nu|$ edge mode(s) on each edge, whose energy (frequency) traverses the gap when $\phi$ varies from $0$ to $2\pi$ \cite{krausPRL2012,krausPRL2012b,poshakinskiyPRL2014}. The sign of gap Chern number determines the chirality (group velocity) of the left edge modes \cite{poshakinskiyPRL2014}. According to the observed topological edge modes in Figs. \ref{incommlongedgestates}(b-d), we can confirm that the bulk-boundary correspondence is valid in our system, and therefore the midgap edge states are indeed topologically protected. In other words, they can be regarded as topological phonon polaritons. We can further validate this correspondence by investigating the minigaps with more midgap states, for instance, the minigap covering $\omega$, which has an IDOS of 0.24 and a Chern number of $\nu=2$, leading to two edge modes on both left and right edges (i.e., four edge modes in total), respectively. Since these midgap edge states in the minigaps will not contribute significantly to radiative heat transfer, we will not study them in detail \cite{wangPRA2021}. Moreover, for the transverse eigenstates, the band gaps are substantially narrower as a consequence of weaker dipole-dipole interactions. This is because transverse eigenstates involve a far-field interaction term that decays very slowly with the distance $r$ as $1/r$ [Eq. (\ref{Gyy})]. Such far-field interactions can result in very long-range hoppings of excited states and effectively reduce the strength of near-field interactions, leading to a reduction in the band gap width. In spite of these differences, the qualitative behavior is still quite similar to the conventional AAH model. Therefore, we will not discuss them in more details in this work.

\begin{figure*}[ht]
	\centering
	\includegraphics[width=1\linewidth]{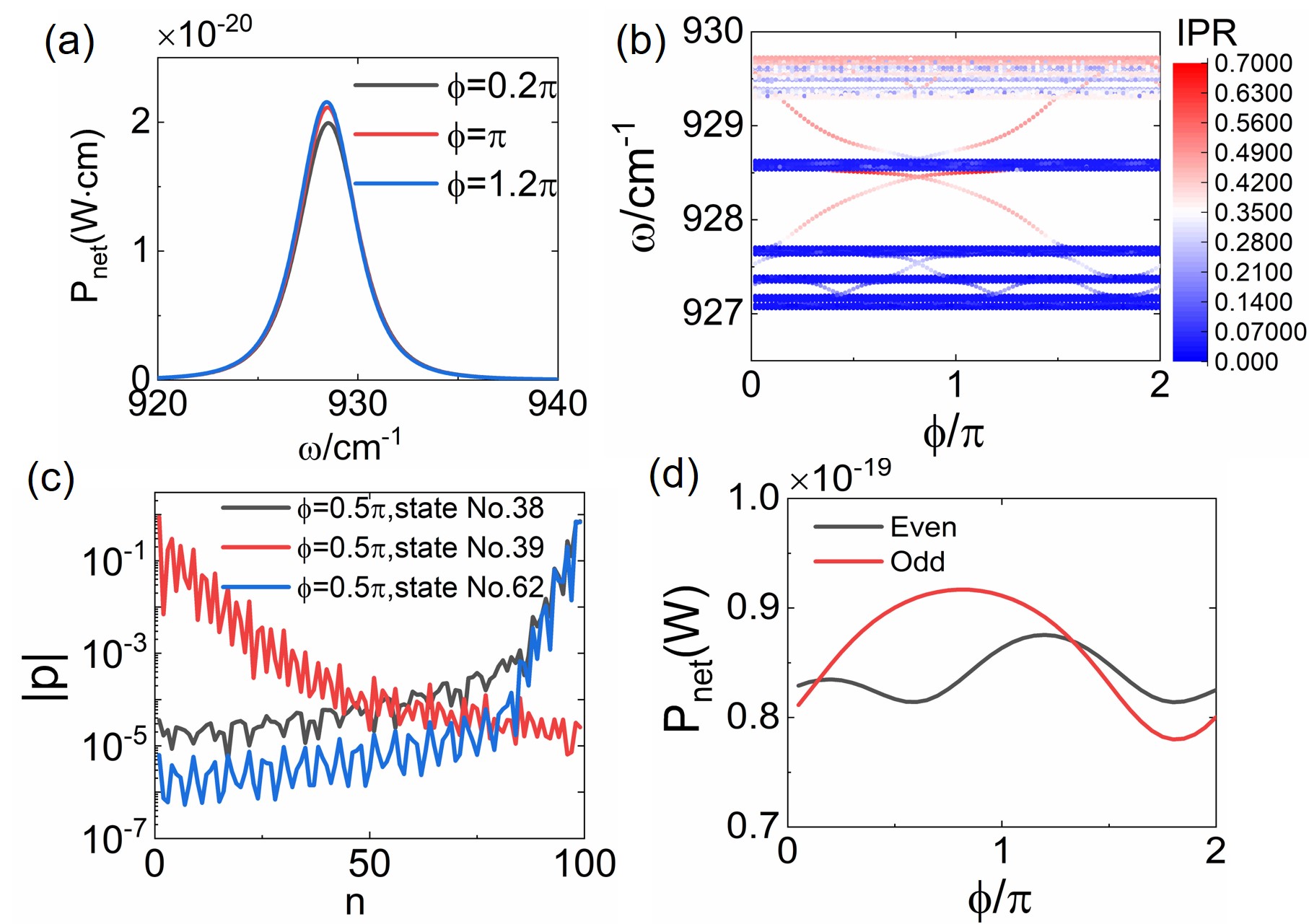}
	\caption{Radiative heat transfer in the incommensurate lattice of $\beta=(\sqrt{5}-1)/2$. (a) Spectral net heat transfer rate for $\phi=0.2\pi$, $\pi$ and $1.2\pi$. (b) Band structure for an array with an odd number of NPs ($N=99$) as a function of modulation phase $\phi$. (c) Dipole moment distributions for the midgap states for $\phi=0.5\pi$. (d) Total net radiative heat rate as a function of modulation phase $\phi$ for both even and odd cases.} 
	\label{incommeta03heat}
\end{figure*}

Now let us consider how TPhPs affect radiative heat transfer in such a system. We assume the first NP has a temperature of $T_1=310~\mathrm{K}$ while all of the other NPs' temperatures are kept as $T_j=300~\mathrm{K}$, $j=2,3,...,N$. This choice of temperature distribution is to partially match the resonance frequency of phonon polaritons in SiC. The net spectral heat rate from the first NP to the last NP, $P^{N1}_\mathrm{net}(\omega)=\mathcal{P}_{1\rightarrow N,\omega}-\mathcal{P}_{N\rightarrow 1,\omega}$ for $\phi=0.2\pi, \pi$ and $1.2\pi$, is plotted in Fig. \ref{incommeta03heat}(a). Note the results take the contributions of both longitudinal and transverse polarizations into account, although we only investigate the longitudinal eigenstate spectra. It is found that there is considerable differences between the maximum spectral radiative heat rate $P^{N1}_\mathrm{net,max}$ in these three cases. For $\phi=0.2\pi$, since there is not any TPhP [Fig. \ref{incommlongedgestates}(a-b)], $P^{N1}_\mathrm{net,max}$ is the smallest. In both cases of $\phi=\pi$ and $1.2\pi$, TPhPs are present in the main gaps [Fig. \ref{incommlongedgestates}(a,c,d)], which enhance the radiative heat transfer process. Moreover, $P^{N1}_\mathrm{net,max}(\phi=1.2\pi)$ is slightly larger than $P^{N1}_\mathrm{net,max}(\phi=\pi)$, which is because in the case of $\phi=1.2\pi$, there are two nearly degenerate eigenstates near the central band edge (also near the phonon polariton resonance frequency $\omega\sim928.5\mathrm{cm}^{-1}$ of a single NP) that can further enhance the long-range heat transfer (considering the length of the chain is around $\sim100d=60\mathrm{\mu m}$).

For the AAH chain, there is a well-known even-odd effect with respect to the number of lattice sites. In Fig. \ref{incommeta03heat}(b), the band structure for an $N=99$ lattice is given, where we can see one of the topological edge modes in the main gap does not change compared to the even case while the other is largely shifted in $\phi$ \cite{guoOL2018}. We can confirm it still fulfills the gap-labeling theorem and the dipole moment distributions of topological edge states are presented in Fig. \ref{incommeta03heat}(c). It is also found that this shift of topological edge mode has a significant effect on radiative heat transfer. Figure \ref{incommeta03heat}(d) shows the total net radiative heat transfer rate (integrated over all frequencies) $P^{N1}_\mathrm{net}=\mathcal{P}_{1\rightarrow N}-\mathcal{P}_{N\rightarrow 1}$ for the even and odd cases as a function of modulation phase $\phi$, in which it is clearly seen that this variation closely resembles the evolution of topological edge modes with $\phi$ in both even and odd cases [Figs. \ref{incommlongedgestates}(a) and \ref{incommeta03heat}(b)]. That is, the existence of TPhPs can significantly enhance $P^{N1}_\mathrm{net}$, and if there are more topological edge states approaching the central band ($\omega\sim928.5\mathrm{cm}^{-1}$, namely near the phonon polariton resonance frequency of a single NP), the larger $P^{N1}_\mathrm{net}$ is. Since the shift of topological mode in the odd case leads to a major change of the $\phi$-dependence compared to the even case, we can further confirm that TPhPs play an important role in this long-range radiative heat transfer process.

\section{Commensurate lattice}
In this section we further discuss the cases of rational $\beta$'s, namely, commensurate lattice. We will mainly focus mainly two cases, $\beta=1/2$ and $\beta=1/4$.
\begin{figure*}[htbp]
	\centering
	\includegraphics[width=1\linewidth]{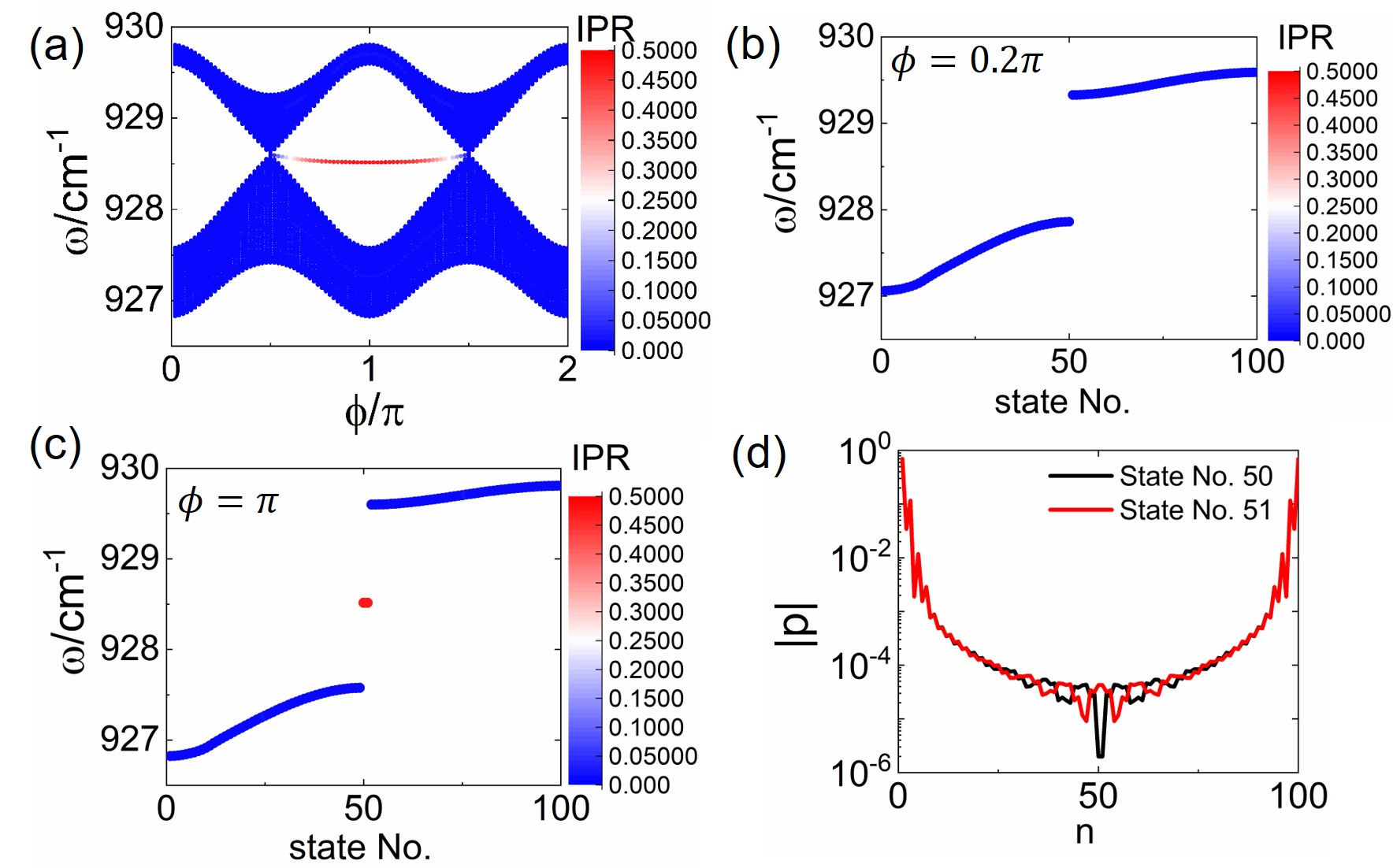}
	\caption{Longitudinal band structures and midgap modes in the case of $\beta=1/2$. (a) Band structure as a function of modulation phase $\phi$. (b) Eigenstate distribution for $\phi=0.2\pi$. (c) Eigenstate distribution for $\phi=\pi$. (d) Dipole moment distribution for the midgap states at $\phi=\pi$.} 
	\label{b1to2longedgestates}
\end{figure*}
\subsection{The case of $\beta=1/2$}

In Fig. \ref{b1to2longedgestates}(a), the eigenstate spectrum for $\beta=1/2$ is presented with $N=100$, $d=0.6\mathrm{\mu m}$ and $\eta=0.3$. There is no band gap in the spectrum, with two degenerate points which can be regarded as Dirac points \cite{lauPRL2015,ganeshanPRL2013}. A quasi-zero-energy mode (that is near the resonance frequency of phonon polaritons) exists and connects the two Dirac points. Unlike the situation of the conventional AAH model under nearest-neighbor (NN) approximations without long-range hoppings, the frequency of this quasi-zero-energy mode is not kept constant with $\phi$ but slightly changes \cite{ganeshanPRL2013,liPRB2014,caoLPL2017}. We also note in the commensurate system, the band is no longer flat and varies substantially with $\phi$, since the translational invariance only exists for discrete values of $\phi$ ($\phi=\pi$ for $\beta=1/2$) \cite{krausPRL2012b}. Although the entire system as a family of 1D lattices does not have a full band gap, when a specific configuration with a fixed $\phi\neq 0.5\pi$ or $\pi$ is considered, there is still a band gap, as shown in Fig. \ref{b1to2longedgestates}(b) for $\phi=0.2\pi$ and Fig. \ref{b1to2longedgestates}(c) for $\phi=\pi$. As expected, there is no midgap state in the case of $\phi=0.2\pi$ while two degenerate midgap states emerge in the band gap for the case of $\phi=\pi$, whose dipole moment distributions are given in Fig. \ref{b1to2longedgestates}(d). These midgap modes are indeed localized over the boundaries. As discussed in previous works for the off-diagonal AAH model with NN hoppings under $\beta=1/2$ \cite{ganeshanPRL2013}, the topology property for such a system, as a family of 1D systems, can be more conveniently understood in a Majorana basis, which can be described by a $Z_2$ topological index. This index is 0 for $\cos\phi>0$ while it is 1 for $\cos\phi<0$ in our system, consistent to the emergence of quasi-zero-energy modes. Note although there are long-range dipole-dipole interactions in the present system, they can be regarded as small perturbations which do not affect the robustness of topological edge modes as implied in Ref. \cite{ganeshanPRL2013}, since the interparticle distance is deep-subwavelength and the nearest-neighbor coupling dominates. On the other hand, for a specific $\phi$, the band topology can be described by the Zak phase for 1D systems like the conventional SSH model with trivial chiral symmetry breaking due to next-nearest-neighbor (NNN) and high-order hoppings, which is well understood in previous works of the authors and others \cite{wangPRB2018b,lingOE2015,pocockArxiv2017}.

\begin{figure*}[htbp]
	\centering
	\includegraphics[width=1\linewidth]{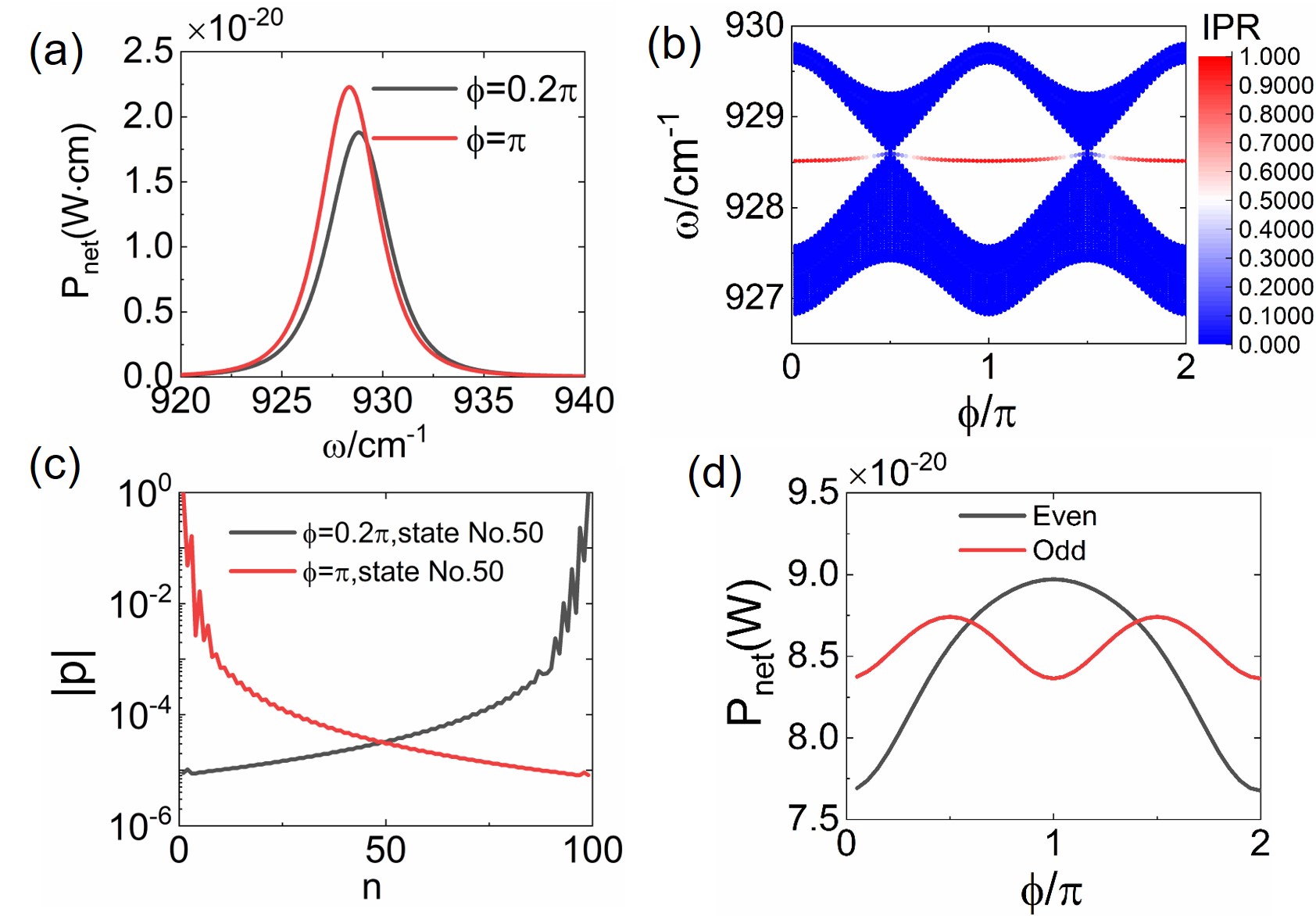}
	\caption{Radiative heat transfer in the case of $\beta=1/2$. (a) Spectral net heat transfer rate for $\phi=0.2\pi$ and $\pi$. (b) Band structure for an array with an odd number of NPs ($N=99$) as a function of modulation phase $\phi$. (c) Dipole moment distributions for the midgap states at $\phi=0.2\pi$ and $\pi$ in the odd case. (d) Total net radiative heat rate as a function of modulation phase $\phi$ for both even and odd cases.} 
	\label{b1to2heat}
\end{figure*}

Let us proceed to the calculation of radiative heat transfer. The net spectral heat rate from the first NP to the last NP, $P^{N1}_\mathrm{net}(\omega)$ for $\phi=0.2\pi$ and $\pi$, is plotted in Fig. \ref{b1to2heat}(a). A significant difference between the spectrum of the topological phase ($\phi=\pi$) and that of the topologically trivial case ($\phi=0.2\pi$) is observed, in which the former shows a considerably larger maximum spectral radiative heat rate $P^{N1}_\mathrm{net,max}$, indicating the enhancement of radiative heat transfer due to TPhPs. Moreover, the spectral position of this maximal value is different between topologically nontrivial and trivial cases. This is because for the former case, $P^{N1}_\mathrm{net,max}$ comes from the excitation of TPhPs in the gap while in the latter case it is due to the transport enhancement originated from the band edges of bulk spectrum \cite{wangPRM2020}. As expected, there is also an even-odd effect for $\beta=1/2$. Figure \ref{b1to2heat}(b) shows the eigenstate spectrum for $N=99$. It is found in this odd case, quasi-zero-energy mode exists for all $\phi$'s. However, we note these edge states only localize over one of the boundaries [Fig. \ref{b1to2heat}(c)], different from the even case in which the edge states are localized over both boundaries (therefore with lower IPRs $\sim0.5$) [Fig. \ref{b1to2longedgestates}(c)]. As discussed in Ref. \cite{ganeshanPRL2013}, in the Majorana basis, there always exists a single Majorana localized on one of the edge sites. 

We compare the total net radiative heat transfer rate $P^{N1}_\mathrm{net}$ for the even and odd cases as a function of modulation phase $\phi$ in Fig. \ref{b1to2heat}(d). In the even case, the variation is closely related to the evolution of topological edge modes with $\phi$ [Figs. \ref{b1to2longedgestates}(a), where the maximum value is resided at $\phi=\pi$ with topological edge states with highest localization degree (i.e., IPR) while the topologically nontrivial regime shows the smallest heat rate. On the other hand, for the odd case, since topological edge states are present for all $\phi$'s, the amplitude of variation of $P^{N1}_\mathrm{net}$ with $\phi$ is small, and the maximum heat transfer rate emerges at the Dirac points. The on-average $P^{N1}_\mathrm{net}$ in the odd case is still large, comparable with the heat transfer rate in the topological phase in the even configuration, while it is considerably lower than the maximum $P^{N1}_\mathrm{net}$ in the even case. Note the length of the array in the odd case is shorter than that of the even case. This important difference derives from the fact that two-sided edge states can show a stronger augmentation of long-range energy transfer than one-side edge states, consistent with previous findings in Ref. \cite{ottPRB2020}.  Therefore, we can further confirm that the presence of TPhPs can indeed enhance long-range heat transfer significantly.

\subsection{The case of $\beta=1/4$}

\begin{figure*}[htbp]
	\centering
	\includegraphics[width=1\linewidth]{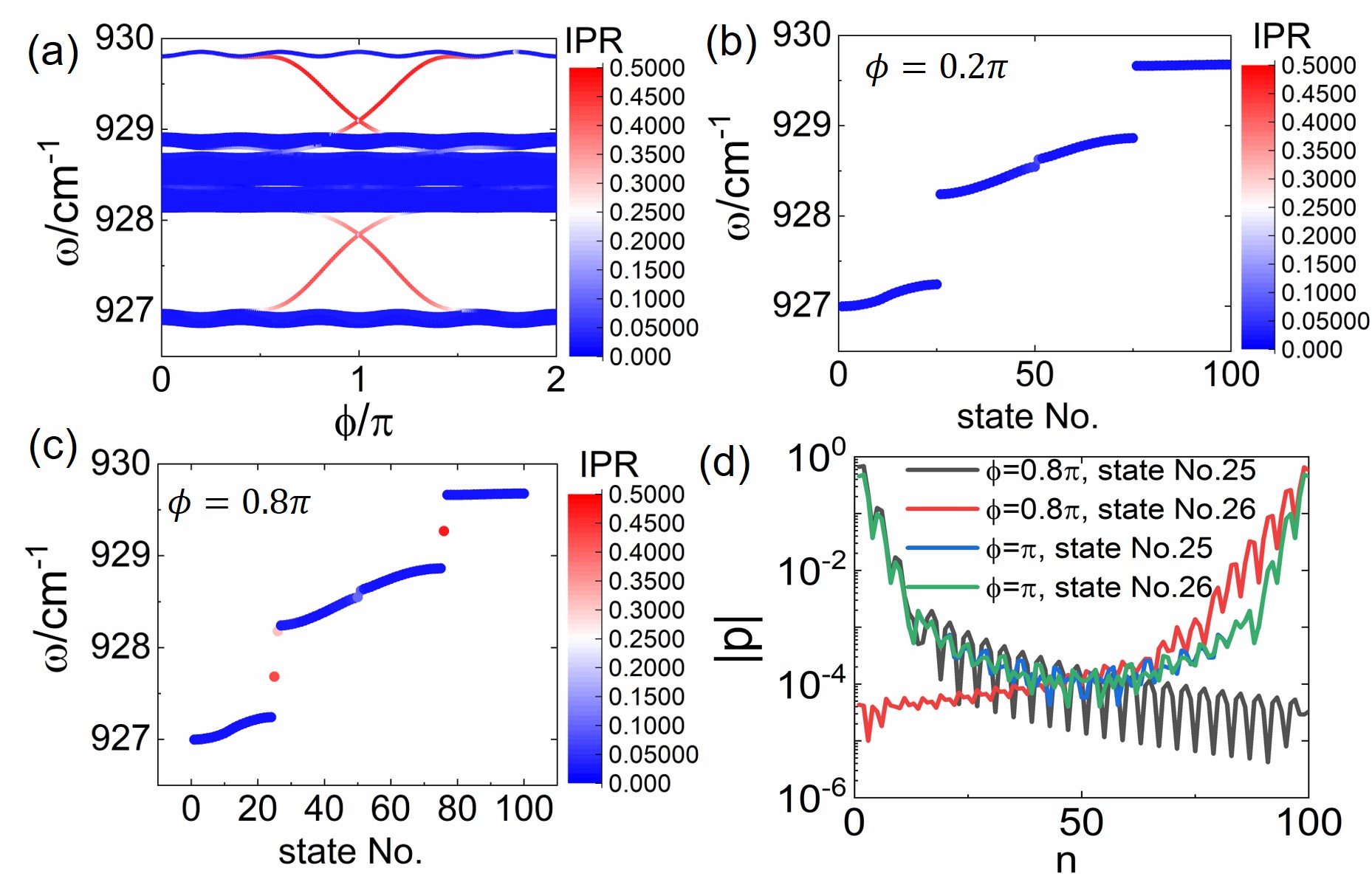}
	\caption{Longitudinal band structures and midgap modes in the case of $\beta=1/4$. (a) Band structure as a function of modulation phase $\phi$. (b) Eigenstate distribution for $\phi=0.5\pi$. (c) Eigenstate distribution for $\phi=\pi$. (d) Dipole moment distributions for the midgap states.} 
	\label{b1to4longedgestates}
\end{figure*}

We further investigate the case of $\beta=1/4$, whose eigenstate spectrum as a function of $\phi$ for a lattice with $N=100$ NPs is shown in Fig. \ref{b1to4longedgestates}(a). Due to this periodic modulation, the band structure breaks into four bands, with two clearly visible main band gaps and one very small gap in the middle of the spectrum. For the conventional AAH model with only NN coupling, there is no such a gap and the two middle bands are connected by four Dirac points \cite{lauPRL2015}. The difference here is due to the NNN and high-order hoppings as discussed in detail in Ref. \cite{ganeshanPRL2013}. We also note in each of the main gaps, there are two midgap modes with high IPRs. These modes are also topologically protected edge modes. Like previous cases, these modes only exist for a specific range of $\phi$. For instance, for the case of $\phi=0.2\pi$ [Fig. \ref{b1to4longedgestates}(b)], there are three band gaps recognized, without any midgap states, while for $\phi=0.8\pi$ [Fig. \ref{b1to4longedgestates}(c)], two midgap states in the lower main gap and one midgap state in the upper main gap are clearly observable. The dipole moment distributions of the two midgap states in the lower main gap (state numbers are 25 and 26) are shown in Fig. \ref{b1to4longedgestates}(d), they are localized over the left and right boundaries respectively. Moreover, for $\phi=\pi$, the two midgap modes cross with each other and the two midgap states become nearly degenerate, and thus they are localized over both of the boundaries of the chain and almost overlap with each other.

\begin{figure*}[htbp]
	\centering
	\includegraphics[width=1\linewidth]{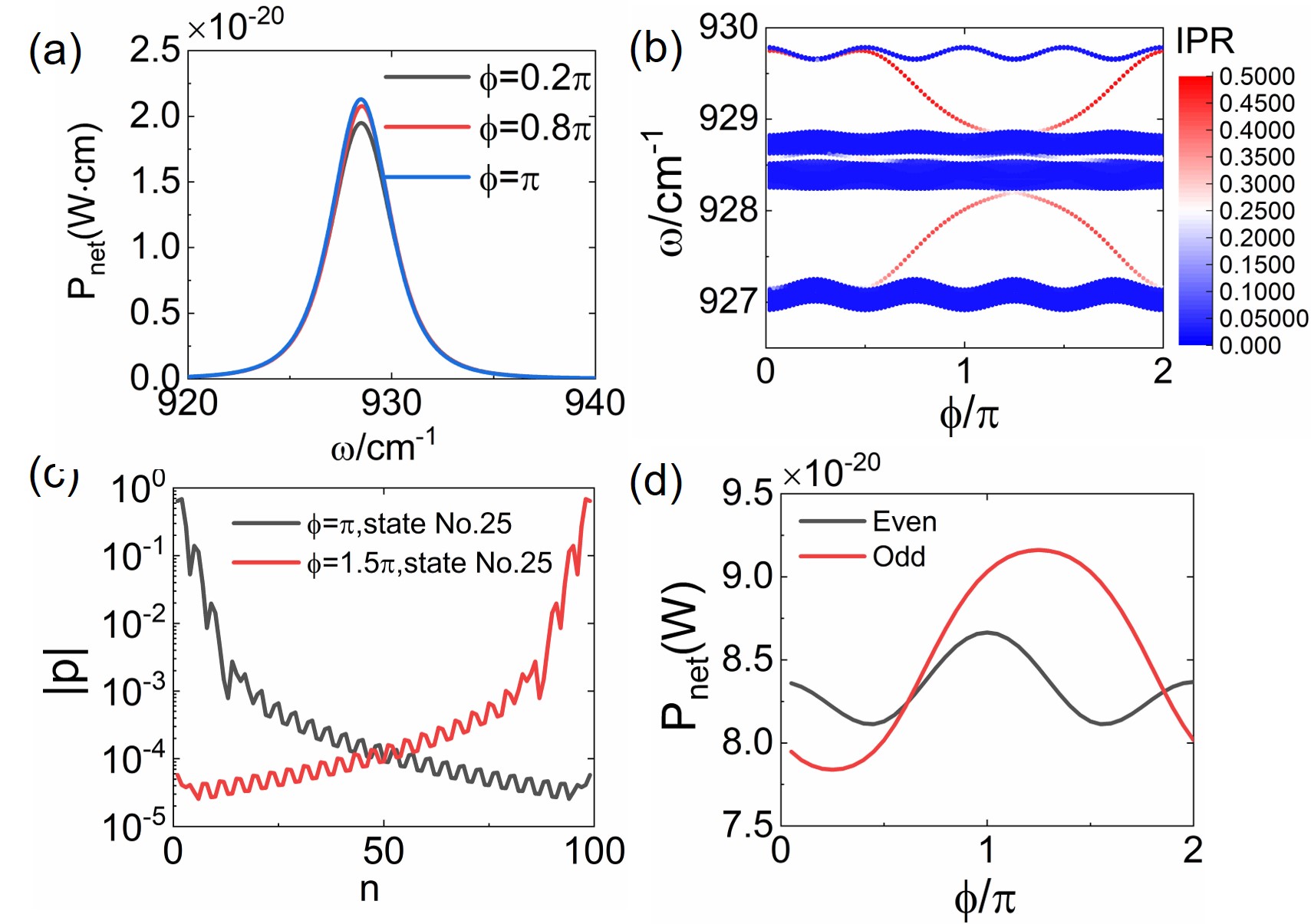}
	\caption{Radiative heat transfer in the case of $\beta=1/4$. (a) Spectral net heat rate for $\phi=0.2\pi$ and $\pi$. (b) Band structure for an array with an odd number of NPs ($N=99$) as a function of modulation phase $\phi$. (c) Dipole moment distributions for the midgap states for $\phi=0.2\pi$ and $\pi$ in the odd case. (d)Total net radiative heat rate as a function of modulation phase $\phi$ for both even and odd cases.} 
	\label{b1to4heat}
\end{figure*}

In this case, the topological properties of the gaps are determined by the Chern number. As aforementioned, the gap-labeling theorem is still valid in the commensurate case but the solution is no longer unique like the incommensurate case. Therefore Chern number for a band $n$ here can only be directly computed by formally carrying out Berry curvature integration through the ``ancestor" 2D quantum Hall model \cite{liPRB2014,poshakinskiyPRL2014}: 
\begin{equation}
\nu_n=\frac{1}{2\pi}\int_{0}^{2\pi}\int_{0}^{2\pi/(2d)}{dkd\phi(\partial A_k/\partial\phi-\partial A_\phi/\partial k)}
\end{equation}
in a 2D parameter space with $\phi$ being the wavenumber in the extended dimension in addition to the wavenumber $k$ in the real 1D space (since the 1D chain is periodic). Here $A_k=i\langle\psi_n(k)|\partial_k|\psi_n(k)\rangle$
and $A_\phi=i\langle\psi_n(k,\phi)|\partial_\phi|\psi_n(k,\phi)\rangle$ with $\psi_n$ being the wavefunction in the band $n$, which can be solved by applying the Bloch theorem under periodic boundary condition and consists of the dipole moments of four NPs in a unit cell in our system. The gap Chern number is the sum of the band Chern numbers below the gap. It is found for the lower gap, $\nu=-1$ while for the upper gap, $\nu=+1$. These Chern numbers guarantee the existence of $|\nu|=1$ topological edge mode at both of the edges in both gaps. Hence the edge modes emerge at the two main gaps are indeed topologically protected.

Figure \ref{b1to4heat}(a) shows the net spectral heat rate from the first NP to the last NP, $P^{N1}_\mathrm{net}(\omega)$ for $\phi=0.2\pi, 0.8\pi$ and $\pi$. Like previous cases, considerable differences between the maximum spectral radiative heat rate $P^{N1}_\mathrm{net,max}$ in these three cases are observed. $P^{N1}_\mathrm{net,max}$ at $\phi=0.2\pi$ is the smallest due to the absence of TPhPs [Fig. \ref{b1to4longedgestates}(a-b)], while in both situations of $\phi=0.8\pi$ and $\pi$, the presence of TPhPs [Fig. \ref{b1to4longedgestates}(a,c)] can substantially enhance the heat transfer rate present in the main gaps. Like previous cases, $P^{N1}_\mathrm{net,max}(\phi=\pi)$ is slightly larger than $P^{N1}_\mathrm{net,max}(\phi=0.8\pi)$,  thanks to the existence of two nearly degenerate topological edge states near central band [Fig. \ref{b1to4longedgestates}(a)] that can further enhance the long-range heat transfer.

The even-odd effect is still persistent, demonstrated by the eigenstate spectrum for $N=99$ shown in Fig. \ref{b1to4heat}(b). It is seen that one of the topological edge modes significantly shifts along the $\phi$ coordinate. The dipole moment distributions of two eigenstates at $\phi=\pi$ and $1.5\pi$ are further presented in Fig. \ref{b1to4heat}(c) to confirm they are topological edge states. The shift of topological modes, as expected, has a significant influence on the $\phi$-dependence of total heat transfer rate $P^{N1}_\mathrm{net}$ [Fig. \ref{b1to4heat}(d)]. The general trend in both even and odd cases follows the evolution of topological edge modes like previous cases. As a result of two degenerate TPhPs approaching closely the central band,  in the odd case (at $\phi\sim1.2\pi$), it exhibits the largest total heat transfer rate. 

\begin{figure*}[htbp]
	\centering
	\includegraphics[width=1\linewidth]{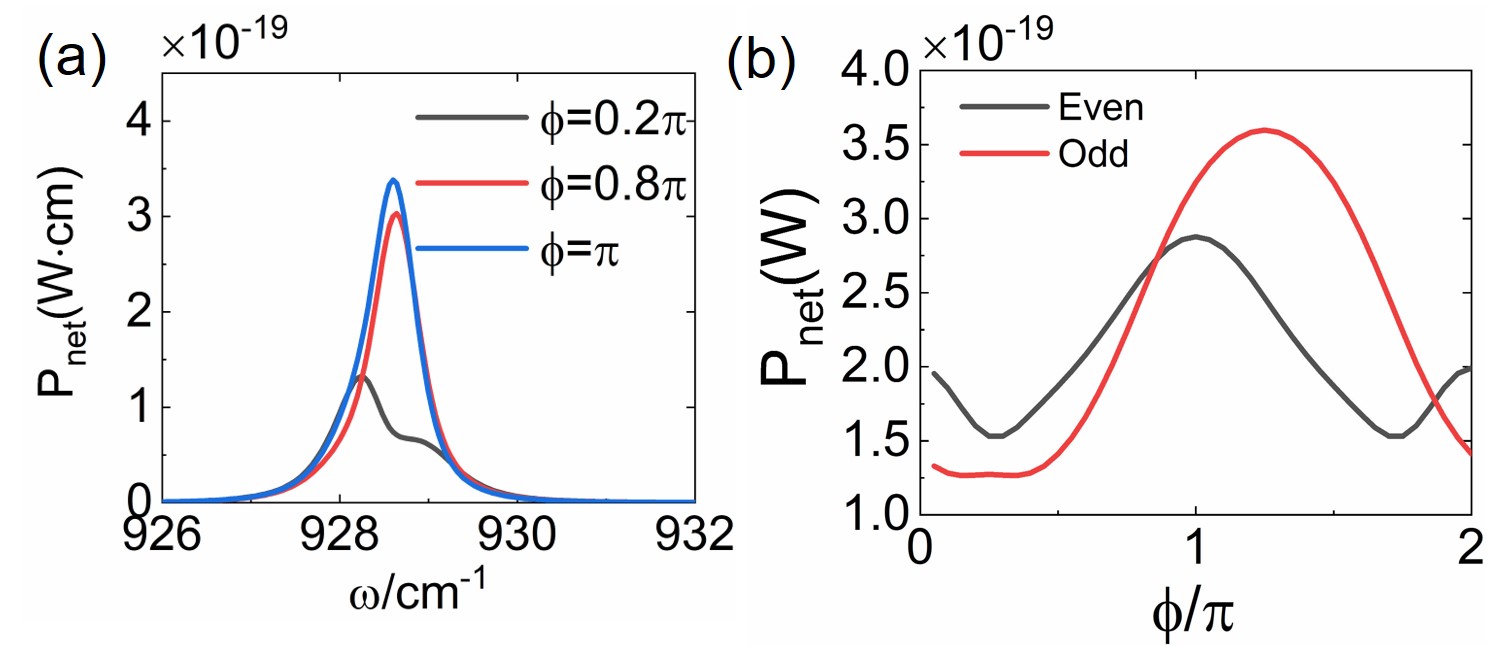}
	\caption{Radiative heat rate in the case of $\beta=1/4$ with an artificial decay rate $\gamma=1\mathrm{cm}^{-1}$. (a) Spectral heat rate for $\phi=0.2\pi$ and $\pi$. (b) Total radiative heat rate as a function of modulation phase $\phi$.} 
	\label{b1to4heatlowloss}
\end{figure*}

We would like to discuss the role of dissipation of the SiC material in the present study of TPhP enhanced radiative heat transfer. We assume a low-loss circumstance of $\gamma=1\mathrm{cm^{-1}}$ for the permittivity in Eq. (\ref{permittivity}) with other parameters unchanged. In this case, the eigenstate spectrum shows almost no difference with the case of $\gamma=5\mathrm{cm^{-1}}$ (not shown here). The net spectral heat transfer rate from the first NP to the last NP $P^{N1}_\mathrm{net}(\omega)$ for $\phi=0.2\pi, 0.8\pi$ and $\pi$ is shown in Fig. \ref{b1to4heatlowloss}(a). More significant differences between these spectra than previous high-loss cases are observed. On one hand, the topological phases ($\phi=0.8\pi$ and $\pi$) show several times larger maximum spectral radiative heat rates $P^{N1}_\mathrm{net,max}$ than the topologically trivial case ($\phi=0.2\pi$), clearly confirming the enhancement of radiative heat transfer due to TPhPs. Moreover, the spectral position of this maximal value is different for the topologically nontrivial and trivial cases, similar to the reason of the difference in Fig. \ref{b1to2heat}(a). That is, for the topologically nontrivial cases, $P^{N1}_\mathrm{net,max}$ is dominated by the excitation of TPhPs in the gap while in the topologically trivial case it is due to the transport mediated by the eigenstates in bulk bands \cite{wangPRM2020}. In Fig. \ref{b1to4heatlowloss}(b), total radiative heat rate $P^{N1}_\mathrm{net}$ as a function of modulation phase $\phi$ for the even and odd lattices is provided, which unequivocally shows the giant enhancement brought by the existence of TPhPs. Notably in the odd case, the presence of TPhPs demonstrates nearly three times larger total radiative heat transfer rate $P^{N1}_\mathrm{net}$ than that of the topologically trivial lattice. Therefore, the damping rate in the SiC material, although which is already very low (compared to typical plasmonic materials), is the key factor that limits the enhancement brought by TPhPs \cite{ottIJHMT2022}. We envision that future works to explore phonon polaritonic materials with lower losses will be critical to the further development of the interface between topological photonic systems and radiative heat transfer, especially when it comes to the requirement of successful experimental observation.

\section{Conclusions}
To summarize, we show TPhPs can be realized in 1D bichromatic SiC nanoparticle chains and they can considerably enhance radiative heat transfer for an array much longer than the wavelength of radiation. The introduction of incommensurate modulations on the interparticle distances of 1D periodic chains leads to a mimicry of the off-diagonal AAH model. The eigenstate spectrum with respect to the modulation phase is calculated, which clearly demonstrates that under this type of modulation the chain supports nontrivial topological modes localized over the boundaries. This is because the present system inherits the topological property of two-dimensional IQH systems despite the presence of long-range dipole-dipole interactions. In this circumstance the gap-labeling theorem and corresponding Chern number can still be used to characterize the features of band gaps and topological edge modes. An exception is for the $\beta=1/2$ case whose topological property can be characterize by a $Z_2$ topological index in the Majorana basis. Moreover, based on many-body radiative heat transfer theory for a set of dipoles, we show the presence of topological gaps and midgap TPhPs can considerably enhance radiative heat transfer for an array much longer than the wavelength of radiation. We show how the modulation phase that acts as the synthetic dimension can tailor the radiative heat transfer rate by inducing or annihilating topological modes. We also find if the damping rate of the SiC material can be reduced, the enhancement of radiative heat transfer due to TPhPs can be significantly larger. These findings therefore provide a fascinating route for tailoring near-field radiative heat transfer based on the concept of topological physics.   

As the concept of synthetic dimension due to the interparticle distance modulation can be further extended to create high-order topological systems such as 4D high-order topological insulator and 4D Chern insulator \cite{chenPRX2021}, we expect more physical insights can be obtained if this concept is introduced to the many-particle system. Moreover, although the present system does not exhibit significant non-Hermiticity for the topological band that may alters the topological properties qualitatively, we envision that the presence of strong non-Hermiticity would have qualitative effects on the topological optical modes mediated radiative heat transfer \cite{longhiPRL2019,zengPRB2020,caiPRB2021a}.

\begin{acknowledgments}
This work is supported by the National Natural Science Foundation of China (Nos. 51906144, 52120105009 and 52090063) and Science and Technology Commission of Shanghai Municipality (Nos. 22ZR1432900 and 20JC1414800).
\end{acknowledgments}

\bibliography{aah_model}
\end{document}